\RequirePackage{fix-cm}
\documentclass[smallextended]{svjour3}       
\smartqed  
\usepackage{graphicx}
\usepackage{bbm}
\usepackage[caption=false]{subfig}
\usepackage{hyperref}

\begin{document}

\title{Modelling urban networks using Variational Autoencoders
}


\author{Kira Kempinska       \and
        Roberto Murcio 
}


\institute{Kira Kempinska \& Roberto Murcio \at
              Centre for Advanced Spatial Analysis (CASA) \\
              University College London\\
              London, UK\\           
              \email{kira.kempinska@ucl.ac.uk} \\
           \and
           Kira Kempinska \at
           Alphamoon Ltd \\
           Wrocław, PL\\
           \email{kira.kempinska@alphamoon.ai}           \\  
}

\date{Received: date / Accepted: date}

\maketitle

\begin{abstract}
A long-standing question for urban and regional planners pertains to the ability to describe urban patterns quantitatively.  Cities' transport infrastructure, particularly street networks, provides an invaluable source of information about the urban patterns generated by peoples' movements and their interactions.  With the increasing availability of street network datasets and the advancements in deep learning methods, we are presented with an unprecedented opportunity to push the frontiers of urban modelling towards more data-driven and accurate models of urban forms.

In this study, we present our initial work on applying deep generative models to urban street network data to create spatially explicit urban models. We based our work on Variational Autoencoders (VAEs) which are deep generative models that have recently gained their popularity due to the ability to generate realistic images. Initial results show that VAEs are capable of capturing key high-level urban network metrics using low-dimensional vectors and generating new urban forms of complexity matching the cities captured in the street network data.

\end{abstract}

\section{Introduction}
\label{intro}

Temporal and spatial patterns of human interactions shape our cities making them unique, but, at the same time, create universal processes that make urban structures comparable to each other. A long-standing effort of urban studies focuses on the creation of quantitative models of the spatial forms of cities that would capture their essential characteristics and enable data-driven comparisons. There have been several attempts at studying urban forms using quantitative methods, typically based on complexity theory or network science \cite{arcaute2016cities,barthelemy2008modeling,murcio2015multifrac,buhl2006topological,cardillo2006structural,masucci2009random,strano2013urban}. The approaches create an abstract representation of an urban form to derive its key quantitative characteristics. Although theoretically robust, the abstractions might often be too simplistic to capture the full breadth and complexity of existing urban structures.

With the increasing availability of urban street network data and the advancements in deep learning methods, we are presented with an unprecedented opportunity to push the frontiers of urban modelling towards more data-driven and accurate urban models. In this study, we present our initial work on applying deep generative models to urban street network data to create spatially explicit models of urban networks. We based our work on Variational Autoencoders (VAEs) trained on images of street networks. VAEs are deep generative models that have recently gained their popularity due to the ability to generate realistic images. VAEs have two fundamental qualities that make them particularly suitable for urban modelling. Firstly, they can condense high dimensional images of urban street networks to a low-dimensional representation which enables quantitative comparisons between urban forms without any prior assumptions. Secondly, VAEs can generate new realistic urban forms that capture the diversity of existing cities. 

In the following sections, we show our experiments based on urban street networks from Open Street Map (OSM). The results indicate that VAE trained on the OSM data is capable of capturing critical high-level urban metrics using low-dimensional vectors. The model can also generate new urban forms of structure matching the cities captured in the OSM dataset. All code and experiments for this study are available at \href{https://github.com/kirakowalska/vae-urban-network}{https://github.com/kirakowalska/vae-urban-network}.

\section{Methodology and dataset}
\label{sec:1}

\subsection{Variational Autoencoder}
\label{sec:2}

Variational Autoencoders (VAEs) have emerged as one of the most popular deep learning techniques for unsupervised learning of complicated data distributions. VAEs are particularly appealing because they compress data into a lower-dimensional representation which can be used for quantitative comparisons and new data generation. VAEs are built on top of standard function approximators (neural networks)  efficiently trained with stochastic gradient descent \cite{kingma2013auto}. VAEs have already been used to generate many kinds of complex data, including handwritten digits, faces, house numbers, and predicting the future from static images. In this work, we apply VAEs to street network images to learn low-dimensional representations of street networks. We use the representations to make quantitative comparisons between urban forms without making any prior assumptions and to generate new realistic urban forms. 

\begin{figure*}
  \includegraphics[width=0.95\textwidth]{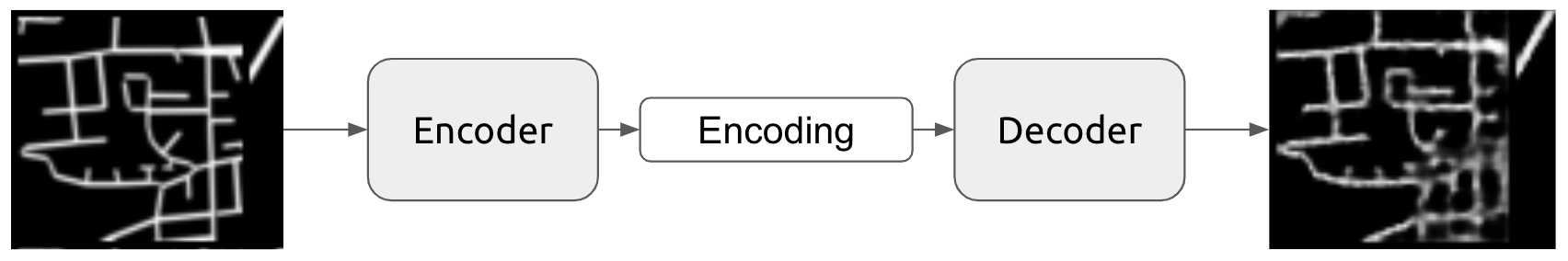}
\caption{Variational Autoencoder takes as input an image of the street network (left), condenses the image to a lower-dimensional encoding (middle) and finally reconstructs the image given the encoding (right).}  
\label{fig:2}       
\end{figure*}

A variational autoencoder consists of an encoder, a decoder, and a loss function. The \textit{encoder} is a neural network. Its input is a datapoint \textit{x}, its output is a hidden representation \textit{z}, and it has weights and biases $\theta$. The goal of the encoder is to 'encode' the data into a latent (hidden) representation space $z$, which has much fewer dimensions that the data. This is typically referred to as a 'bottleneck' because the encoder must learn an efficient compression of the data into this lower-dimensional space. The encoder is denoted by $q_\phi (z|x)$.

The \textit{decoder} is another neural network. Its input is the representation $z$, it outputs a data point $x$, and has weights and biases $\phi$. The decoder is denoted by $p_\phi(x|z)$. The decoder 'decodes' the low-dimensional latent representation $z$ into the datapoint $x$. Information is lost in the process because the decoder translates from a smaller to a larger dimensionality. How much information is lost? The information loss is measured using the reconstruction log-likelihood $\log p_\phi(x|z)$. The measure indicates how effectively the decoder has learned to reconstruct an input image $x$ given its latent representation $z$.

The \textit{loss function} of the variational autoencoder is the sum of the reconstruction loss, given by the negative log-likelihood, and a regularizer. The total loss is the sum of losses $\sum_{i=1}^N l_i$ for $N$ datapoints, where the loss function $l_i$ for datapoint $x_i$ is:

\begin{equation}\label{eq:loss}
l_i (\theta, \phi) = - \mathbbm{E}_{z \sim q_\theta (z|x_i)} [\log p_{\phi} (x_i | z)] + \mathbbm{KL}(q_\theta (z|x_i) || p(z))
\end{equation}

The first term is the reconstruction loss or expected negative log-likelihood of the \textit{i}-th data point. This term encourages the decoder to learn to reconstruct the data. Poor reconstruction of the data $x$ from its latent representation $z$ will incur a large cost in this loss term. The second term is a regularizer that we introduce to ensure that the distribution of the latent values $z$ approaches the prior distribution $p(z)$ specified as a Normal distribution with mean zero and variance one. The regularizer is the Kullback-Leibler divergence between the encoder's distribution $q_\theta (z|x)$ and $p(z)$. It measures how close $q$ is to $p$. The regularizer ensures that the representations $z$ of each data point are sufficiently diverse and distributed approximately according to a normal distribution, from which we can easily sample.

The variational autoencoder is trained using gradient descent to optimize the loss with respect to the parameters of the encoder and decoder $\theta$ and $\phi$. 

\begin{figure} 
  \includegraphics[width=\textwidth]{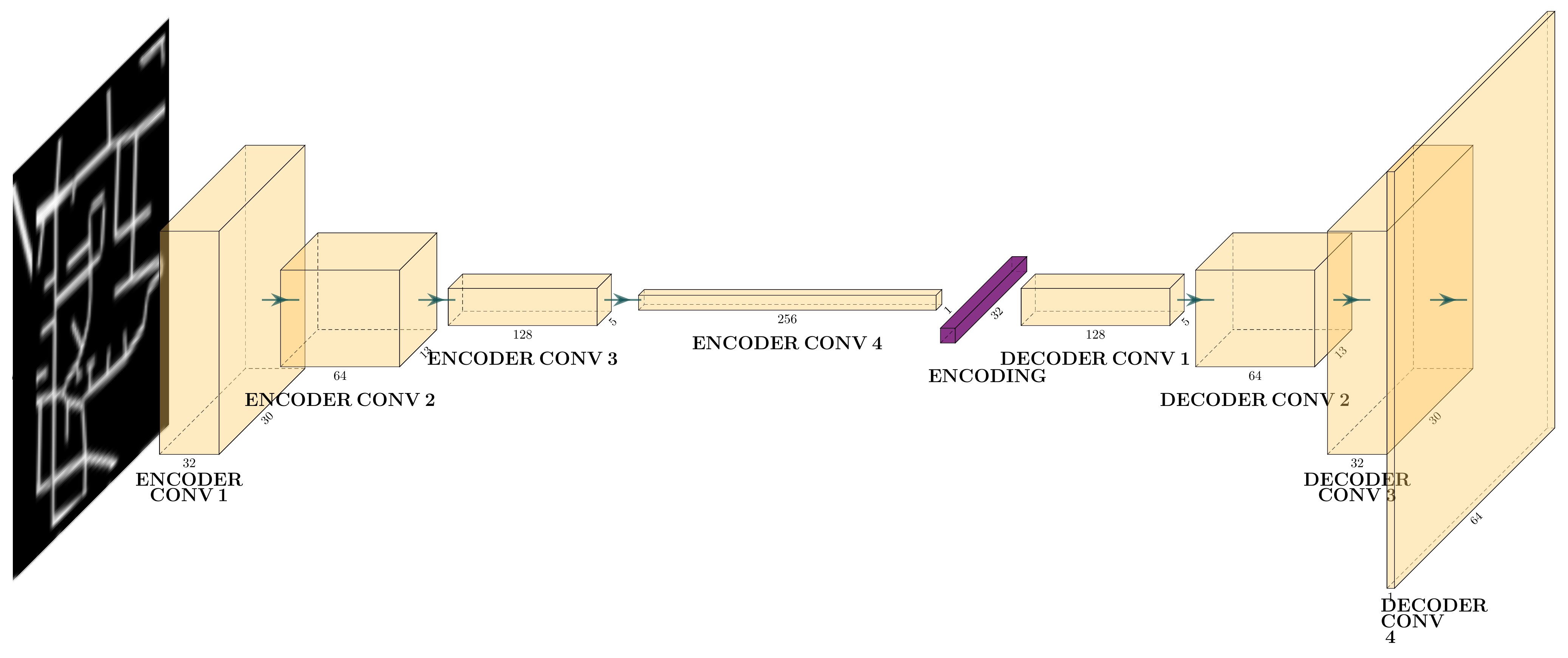}
\caption{Variational autoencoder architecture. Yellow blocks represent convolutional blocks (convolutional layer followed by ReLU layer) with dimensions corresponding to their output dimensions. The purple block is the learnt embedding $z$. }  
\label{fig:vae_architecture_}
\end{figure}

In our work, we selected Convolutional Neural Networks (CNNs) \cite{fukushima1980neocognitron,lecun1990handwritten} as the encoder and decoder architectures. CNNs are deep learning architectures that are particularly well-suited to image data \cite{lecun1995convolutional,krizhevsky2012imagenet} as they consider the two-dimensional structure of images and scale well to high-dimensional images. We tested several CNN architectures and finally chose a network architecture in Figure~\ref{fig:vae_architecture_} with the encoder and the decoder architectures consisting of four convolutional blocks, each with a convolutional and a rectified linear unit (ReLU) layer (which introduces non-linearity to the network). The architecture takes as input an image of size 64 x 64 pixels, convolves the image through the encoder network and then condenses it to a 32-dimensional latent representation. The decoder then reconstructs the original image from the condensed latent representation. We implemented the variational autoencoder using PyTorch library for Python.



\subsection{Street Network Data}

\begin{figure} 
\begin{tabular}{cccccccccc}
  \includegraphics[width=0.08\textwidth]{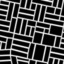} &   \includegraphics[width=0.08\textwidth]{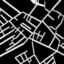} &   \includegraphics[width=0.08\textwidth]{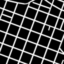} &   
 \includegraphics[width=0.08\textwidth]{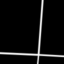}  &   \includegraphics[width=0.08\textwidth]{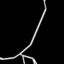} &   \includegraphics[width=0.08\textwidth]{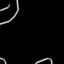} &   \includegraphics[width=0.08\textwidth]{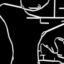} &   \includegraphics[width=0.08\textwidth]{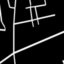} &   \includegraphics[width=0.08\textwidth]{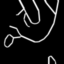} &   \includegraphics[width=0.08\textwidth]{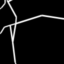} \\
  \includegraphics[width=0.08\textwidth]{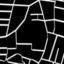} &   \includegraphics[width=0.08\textwidth]{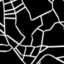} &   \includegraphics[width=0.08\textwidth]{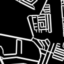} &   \includegraphics[width=0.08\textwidth]{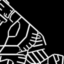}  &   \includegraphics[width=0.08\textwidth]{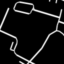} &   \includegraphics[width=0.08\textwidth]{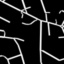} &   \includegraphics[width=0.08\textwidth]{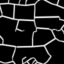} &   \includegraphics[width=0.08\textwidth]{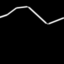} &   \includegraphics[width=0.08\textwidth]{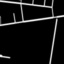} &   \includegraphics[width=0.08\textwidth]{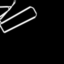} \\
  \includegraphics[width=0.08\textwidth]{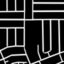} &   \includegraphics[width=0.08\textwidth]{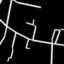} &   \includegraphics[width=0.08\textwidth]{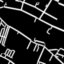} &   \includegraphics[width=0.08\textwidth]{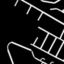}  &   \includegraphics[width=0.08\textwidth]{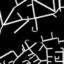} &   \includegraphics[width=0.08\textwidth]{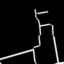} &   \includegraphics[width=0.08\textwidth]{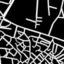} &   \includegraphics[width=0.08\textwidth]{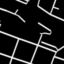} &   \includegraphics[width=0.08\textwidth]{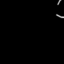} &   \includegraphics[width=0.08\textwidth]{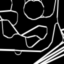} \\
  \includegraphics[width=0.08\textwidth]{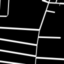} &   \includegraphics[width=0.08\textwidth]{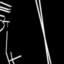} &   \includegraphics[width=0.08\textwidth]{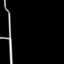} &  \includegraphics[width=0.08\textwidth]{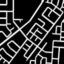}  &   \includegraphics[width=0.08\textwidth]{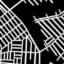} &   \includegraphics[width=0.08\textwidth]{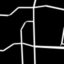} &   \includegraphics[width=0.08\textwidth]{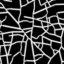} &   \includegraphics[width=0.08\textwidth]{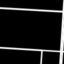} &   \includegraphics[width=0.08\textwidth]{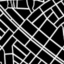} &   \includegraphics[width=0.08\textwidth]{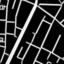} \\
  \includegraphics[width=0.08\textwidth]{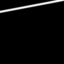} &   \includegraphics[width=0.08\textwidth]{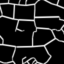} &   \includegraphics[width=0.08\textwidth]{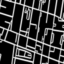} &   \includegraphics[width=0.08\textwidth]{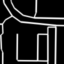}  &   \includegraphics[width=0.08\textwidth]{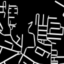} &   \includegraphics[width=0.08\textwidth]{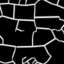} &   \includegraphics[width=0.08\textwidth]{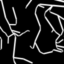} &   \includegraphics[width=0.08\textwidth]{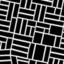} &   \includegraphics[width=0.08\textwidth]{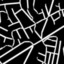} &   \includegraphics[width=0.08\textwidth]{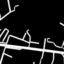} \\
\end{tabular}
\caption{Example images of the street network in randomly selected cities, shown as a square window of 3 x 3\textit{km} centered on the city centre.}
\label{fig:samples}
\end{figure}

The street networks used for model training and testing were obtained from OpenStreetMap \cite{haklay2008openstreetmap} by ranking world cities by 2017 population and then selecting the ones with more than 500,000 inhabitants, for a total of 1059 cities\footnote{We compiled the list of cities from the UN data website \href{http://data.un.org}{http://data.un.org} (accessed December 2018) }.  We saved the street networks as images and, as the Variational autoencoders required images to have a fixed spatial scale,  we extracted a 3 x 3\textit{km} sample from the centre of each city image and resized it to a 64 x 64 pixels binary image. The final dataset contained 1,059 binary images of 64 x 64 pixels, which we split into 80\% training and 20\% testing datasets. During model training, we augmented the training dataset by randomly cropping and flipping the images horizontally. Figure~\ref{fig:samples} shows images for randomly selected cities.

\section{Results}

 \subsection{Reconstruction quality}
\begin{figure*}
  \includegraphics[width=0.95\textwidth]{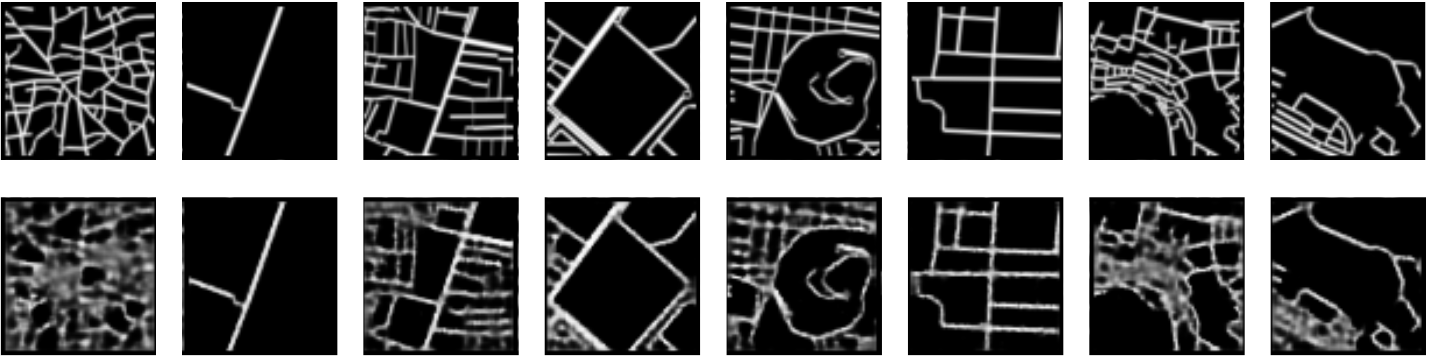}
\caption{Street network reconstructed (bottom) from the original images (top) using the trained autoencoder.}  
\label{fig:reconstruction}       
\end{figure*}

The variational autoencoder was trained to minimise the loss function defined in (\ref{eq:loss}). The training is equivalent to minimising the image reconstruction loss, subject to a regularizer. We can inspect the training quality by visually comparing reconstructed images to their original counterparts. Figure~\ref{fig:reconstruction} shows several examples of reconstructed images of urban street networks. As observed in the examples, the trained autoencoder performs well at reconstructing the overall shape of road networks and their main roads. The quality of the reconstruction drops for very dense road networks when only the overall network shape is captured by the autoencoder (see the leftmost image in Figure~\ref{fig:reconstruction}). The observation suggests that variational autoencoders are better suited for reconstructing images with wide patches of pixels with similar properties rather than narrow stretches such as roads.

\subsection{Urban networks comparison} \label{sec:comparison}
 
The trained autoencoder learnt mapping from the space of street network images (64 x 64 or 4,096 dimensions) to a lower dimensional latent space (32 dimensions). The latent representation stores all the information required to reconstruct the original image of the street network, so it is effectively a condensed representation of the street network that preserves all its connectivity and spatial information. In the lack of well-defined similarity metrics of urban networks, this paper uses the condensed representations as vectors of street network features. Hereafter, we call the vectors \textit{urban network vectors}. Urban network vectors can be used to measure the similarity between different street network forms and to perform further similarity analysis, such as clustering.

\paragraph{Similarity analysis} Firstly, we demonstrated the use of urban network vectors for measuring similarity between urban street forms. We measured the similarity between pairs of vectors as the Euclidean distance. Given two urban network vectors $p=(p_1,p_2,...,p_n)$ and $q=(q_1,q_2,...,q_n)$, where $n=32$ is the size of the latent space $z$, the Euclidean distance between $p$ and $q$ is defined as:
\begin{equation}
d(p,q)=d(q,p)=\sqrt{(q_1 - p_1)^2+(q_2 - p_2)^2+...+(q_n - p_n)^2}.
\end{equation}

\begin{figure} 
\begin{tabular}{ccccccc}
 \includegraphics[width=0.11\textwidth]{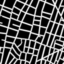} &  \includegraphics[width=0.11\textwidth]{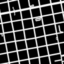} &  \includegraphics[width=0.11\textwidth]{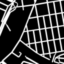} & \includegraphics[width=0.11\textwidth]{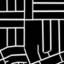} & \includegraphics[width=0.11\textwidth]{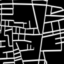} & \includegraphics[width=0.11\textwidth]{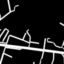} & \includegraphics[width=0.11\textwidth]{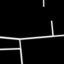} \\
\includegraphics[width=0.11\textwidth]{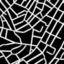} & \includegraphics[width=0.11\textwidth]{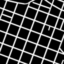} & \includegraphics[width=0.11\textwidth]{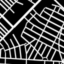} & \includegraphics[width=0.11\textwidth]{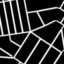} & \includegraphics[width=0.11\textwidth]{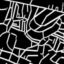} & \includegraphics[width=0.11\textwidth]{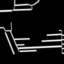} & \includegraphics[width=0.11\textwidth]{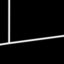} \\
 \includegraphics[width=0.11\textwidth]{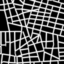} & \includegraphics[width=0.11\textwidth]{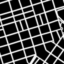} & \includegraphics[width=0.11\textwidth]{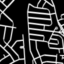} & \includegraphics[width=0.11\textwidth]{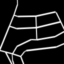} & \includegraphics[width=0.11\textwidth]{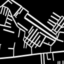} & \includegraphics[width=0.11\textwidth]{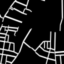} &  \includegraphics[width=0.11\textwidth]{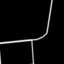} \\
 \includegraphics[width=0.11\textwidth]{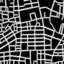}  & \includegraphics[width=0.11\textwidth]{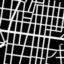}  &  \includegraphics[width=0.11\textwidth]{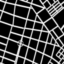}  &  \includegraphics[width=0.11\textwidth]{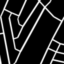}  &  \includegraphics[width=0.11\textwidth]{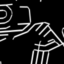}  & \includegraphics[width=0.11\textwidth]{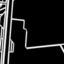}  &  \includegraphics[width=0.11\textwidth]{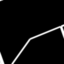}  \\
 \includegraphics[width=0.11\textwidth]{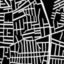} & \includegraphics[width=0.11\textwidth]{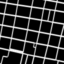} & \includegraphics[width=0.11\textwidth]{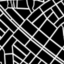} & \includegraphics[width=0.11\textwidth]{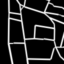} & \includegraphics[width=0.11\textwidth]{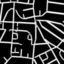} & \includegraphics[width=0.11\textwidth]{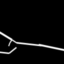} & \includegraphics[width=0.11\textwidth]{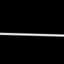}
\end{tabular}
\caption{Street network images (top row) with most similar street networks (rows below) based on the Euclidean distance between their urban network vectors. The latent representations, obtained using the trained encoder, seem to capture well network properties such as density, orientation or road shape.}
\label{fig:similarity}
\end{figure}

Figure~\ref{fig:similarity} shows randomly chosen street networks (top row) and their most similar networks based on the Euclidean distance between their urban street networks. As shown in the figure, the proposed methodology enables finding street networks with matching properties, such as network density, spatial structure and orientation without explicitly including any of the properties in the similarity computation.

\begin{figure*} 
\subfloat[three clusters]{\includegraphics[width = \textwidth]{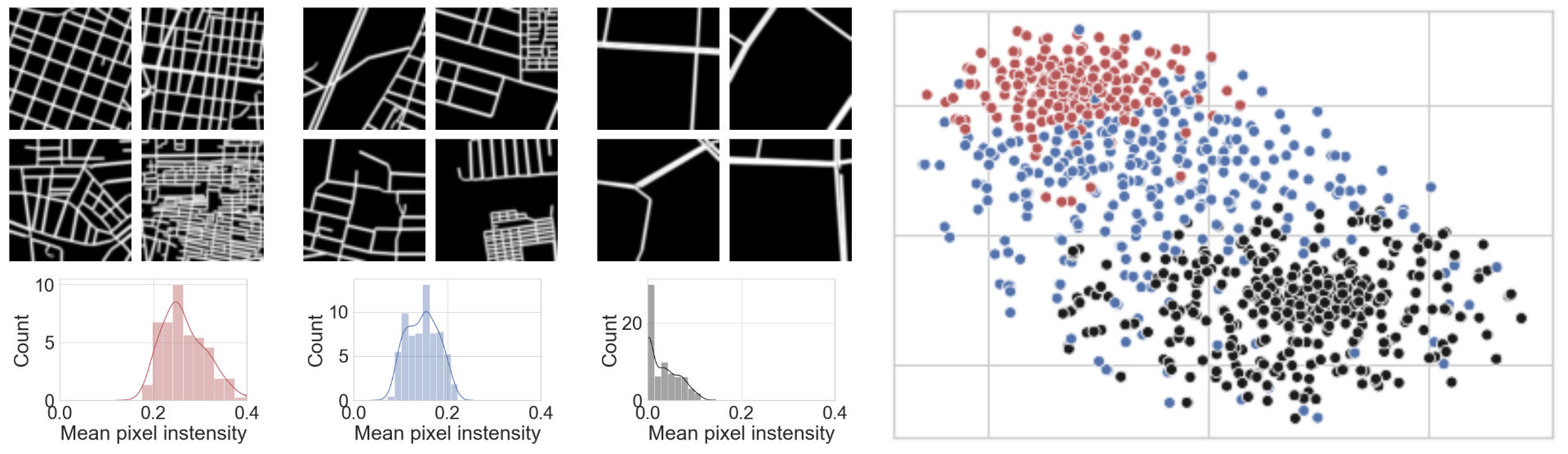}\label{fig:clusters3}}\\
\subfloat[six clusters]{\includegraphics[width = \textwidth]{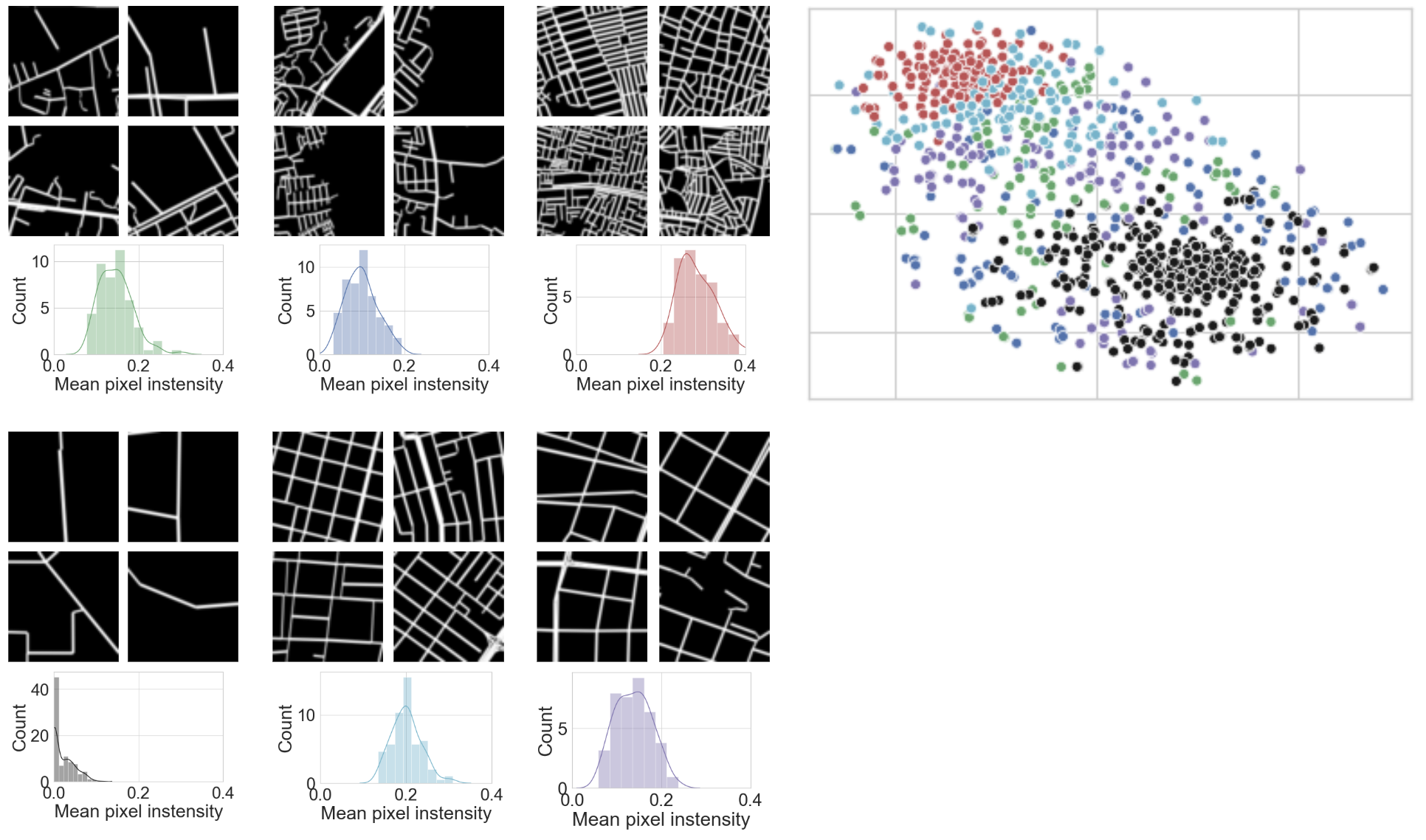}\label{fig:clusters6}}
\caption{(a) Three or (b) six clusters of urban street forms obtained by applying \textit{K-means algorithm} to the condensed urban network vectors. Subfigures show example street networks in each cluster (top left), street network density in each cluster (bottom left) approximated using pixel intensity of street images, and a two-dimensional visualisation of all urban vectors with colour-coded cluster membership.} \label{fig:clustering_}
\end{figure*}

\begin{figure*} 
\includegraphics[width = \textwidth]{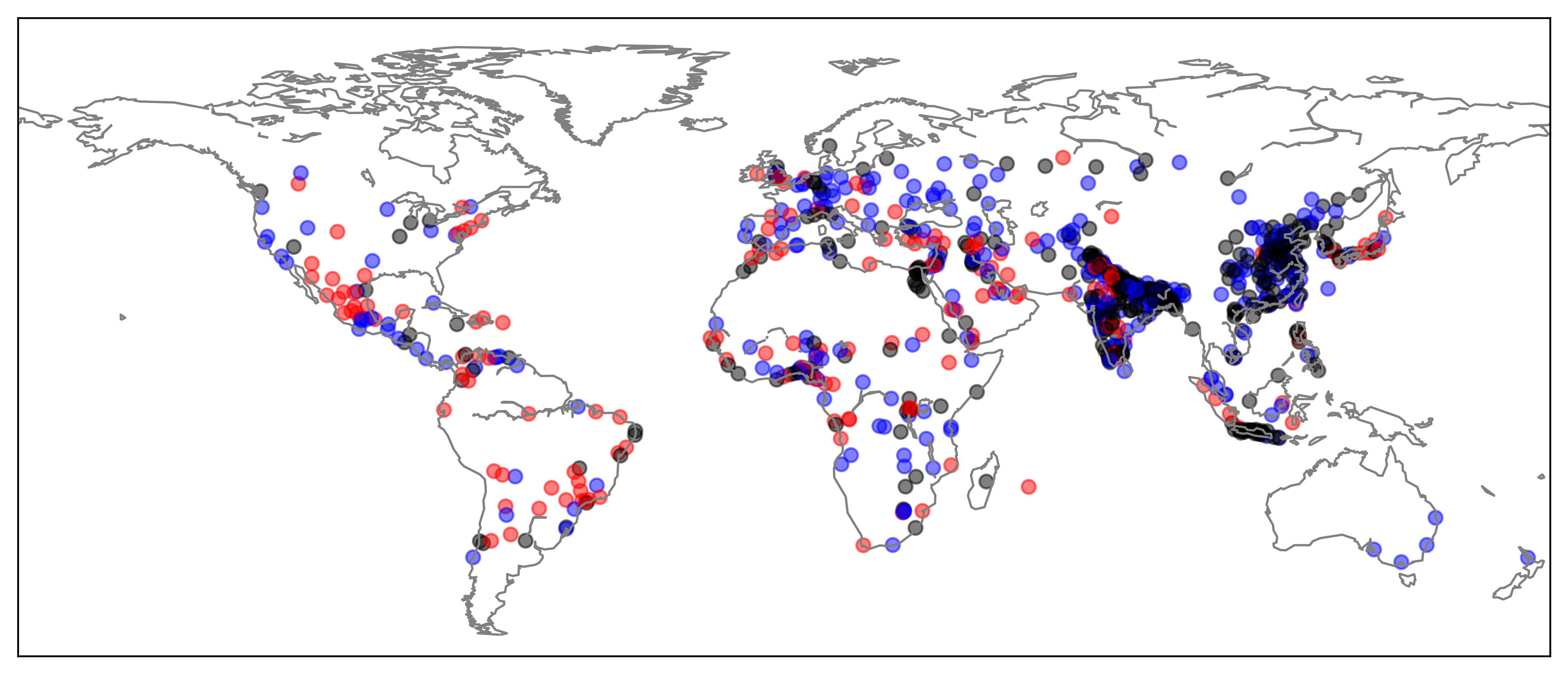}
\caption{Distribution of urban street forms across the globe. Each dot represents a city and is colour-coded according to cluster memberships in Figure~\ref{fig:clusters3}. Despite limited data size, spatial trends start to emerge, such as the concentration of high-density urban networks in California, USA (red cluster) and low-density urban networks in south-eastern Asia (black cluster).} \label{fig:clustering_map}
\end{figure*}

\paragraph{Clustering}
Secondly, we used the urban network vectors to detect clusters of similar urban street forms. We used the \textit{K-means} clustering algorithm \cite{witten2016data}. It is a popular clustering approach that assigns data points to $K$ clusters based on distances to cluster centroids. The algorithm requires specifying the number of clusters $K$ a priori. We identified $K=3$ as the optimal number of clusters for the street image data using the elbow method \cite{dangeti2017statistics}. As shown in Figure~\ref{fig:clusters3}, the obtained clusters seem to separate street networks based on their density only, failing to reflect more subtle network differences, such as road connectivity or road shapes. When we increased the number of clusters to $K=6$ in Figure~\ref{fig:clusters6}, we could differentiate road networks based on more subtle network characteristics, such as disconnectedness of roads in the first cluster (top-left in Figure~\ref{fig:clusters6}) or large gaps in road provision in the second cluster (top-centre in Figure~\ref{fig:clusters6}). We visualised both cluster assignments in Figure~\ref{fig:clustering_} (right) by projecting the thirty-two-dimensional urban network vectors to a two-dimensional grid using T-SNE algorithm \cite{maaten2008visualizing} for dimensionality reduction. The visualisations shows that street networks naturally cluster into three groups that were detected by the $K$-means algorithm. The three clusters are further mapped in Figure~\ref{fig:clustering_map} to investigate spatial patterns in urban form variation.

\subsection{Urban networks generation}

In Section~\ref{sec:comparison}, we used the autoencoder to compress real street images to low-dimensional vectors which we then used to make quantitative comparisons. This employed one strength of variational autoencoders: the ability to \textit{encode} high-dimensional observations as meaningful low-dimensional representations. The second strength pertains to the ability to \textit{generate} realistic urban street forms that match the complexity of urban forms across the globe. The ability could potentially advance the current state-of-the-art in simulations of urban forms and socio-economic processes taking place on urban networks.

\begin{figure*}
  \includegraphics[width=0.95\textwidth]{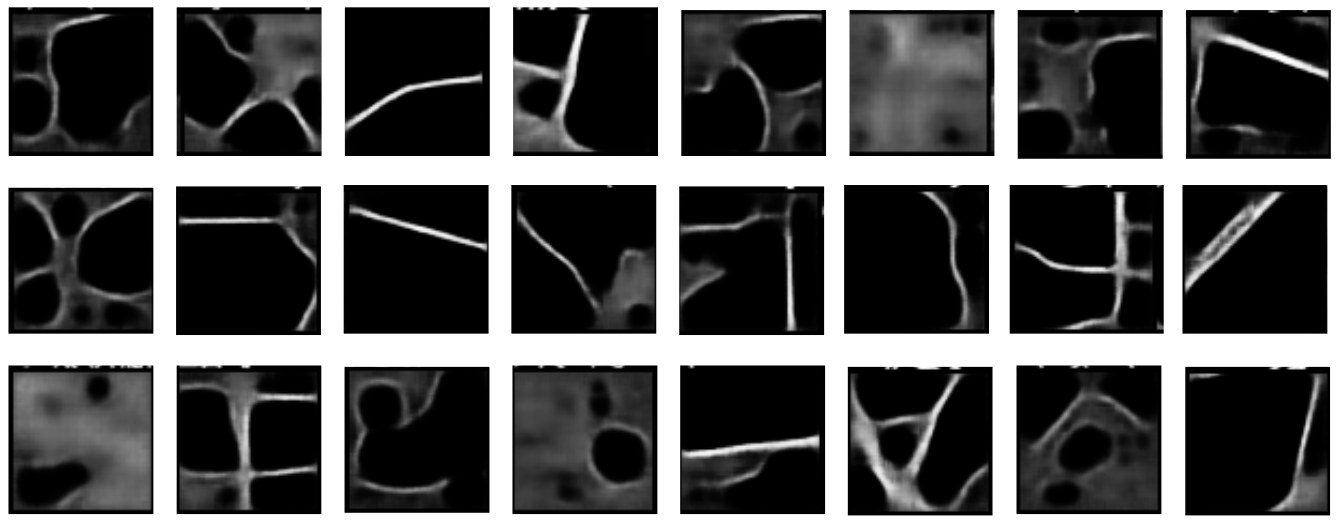}
\caption{Examples of synthetic urban street forms generated by passing a randomly sampled latent code $z$ through the decoder network.}  
\label{fig:generation}       
\end{figure*}

To generate a synthetic urban network, we firstly sample an embedding value $z$ from the prior distribution $p(z)$ specified as a standard Gaussian (see Section \ref{sec:2}) and then pass the value through the decoder network to obtain a corresponding image. Images corresponding to several embedding samples are shown in Figure~\ref{fig:generation}. As shown in the figure, the generated images lack the detail of real street images in Figure~\ref{fig:samples}. Although the samples follow the general structure of road networks with major roads and areas of mixed-density minor roads, the decoder fails to reconstruct details of dense road segments and instead represents them blurred. The problem must be accredited to too few images used in the study. Although the proposed model is flexible enough to model urban street networks, which is confirmed by high-quality reconstructions of real images in Figure~\ref{fig:reconstruction}, it does not see enough images to learn to interpolate between them to sample new forms of street networks to sufficient detail.

\section{Discussion and conclusions}

This study is an early exploration of how modern generative machine learning models such as variational autoencoders could augment our ability to model urban forms. With the ability to extract key urban features from high-dimensional urban imagery, variational autoencoders open new avenues to integrating high-dimensional data streams in urban modelling. The study considered images of street networks, but the proposed methodology could be equally applied to other image data, such as urban satellite imagery.

Variational autoencoders were selected among deep generative models \cite{moosavi2017urban,albert2018modeling} due to their two capabilities: firstly to condense images to low-dimensional representations, secondly to generate new previously unseen images that match the complexity of observed images. The first capability enabled us to extract key urban metrics from street network images, the second gave us the power to generate realistic images of previously unseen urban networks. 

Our results, based on 1,059 city images across the globe, showed that VAEs successfully condensed urban images into low-dimensional urban network vectors. This enabled quantitative similarity analysis between urban forms, such as clustering. What is more, VAEs managed to generate new urban forms with complexity matching that of the observed data. Unfortunately, the resolution of the generated images was low which was accredited to the small size of the dataset. Future work will repeat model training on a much larger corpus of images to improve the generative quality. 

Despite the promising results, the study opens essential questions for future work. The first question pertains to the black-box nature of deep learning models that lack comprehensive human interpretability. This limitation is already receiving much attention in the deep learning literature \cite{lime,shrikumar2017learning,lundberg2017unified}. In this study, the limitation manifests itself in our lack of understanding of how latent space representations of urban networks relate to established network metrics \cite{newman2010networks}. A related question refers to the ability to evaluate the quality of model outputs, i.e. latent representations and synthetic images. Again, quality assessment of deep generative models is a hot topic in the broader deep learning research community (see for example \cite{wu2016quantitative}). Future work could address the problem from the perspective of urban network science.

\section{Declarations}

\subsubsection*{Availability of data and materials}
All data and program source code described in this article is available to any interested parties. The source code and experiments are available at GitHub at the following URL: \href{https://github.com/kirakowalska/vae-urban-network}{https://github.com/kirakowalska/vae-urban-network}. The raw data and datasets generated during this study are available upon request.
\subsubsection*{Competing interests}
The authors declare that they have no competing interests.

\subsubsection*{Funding}
There is no specific funding received for the study.

\subsubsection*{Authors' contributions}
KK designed and implemented the methodology, executed the computer runs, and wrote the initial version of the article. RM prepared street network data and extensively revised the article. Both authors read and approved the final manuscript. 

\subsubsection*{Acknowledgements}
The authors would like to thank Szymon Zareba and Adam Gonczarek (Alphamoon Ltd) for advice on deep generative models during the course of the project. 

\subsubsection*{Authors' information}
KK is a lecturer in geospatial machine learning at the Bartlett's Centre for Advanced Spatial Analysis, University College London, UK and a machine learning researcher at Alphamoon, PL. She develops machine learning algorithms for urban modelling and sensor data mining. Her research interests include geospatial data mining, sensor data fusion and machine learning for sensor networks.

RM is a senior research fellow at the Bartlett's Centre for Advanced Spatial Analysis, University College London, UK. His academic interests include urban complex networks, information transfer in social systems, spatial interaction models and pedestrian flows. One of his main research topics is the application of multifractal measures to different urban aspects, such as street networks and social inequality.  


\bibliographystyle{spmpsci}      
\bibliography{bibliography}   

\begin{thebibliography}{10}
\providecommand{\url}[1]{{#1}}
\providecommand{\urlprefix}{URL }
\expandafter\ifx\csname urlstyle\endcsname\relax
  \providecommand{\doi}[1]{DOI~\discretionary{}{}{}#1}\else
  \providecommand{\doi}{DOI~\discretionary{}{}{}\begingroup
  \urlstyle{rm}\Url}\fi

\bibitem{albert2018modeling}
Albert, A., Strano, E., Kaur, J., Gonz{\'a}lez, M.: Modeling urbanization
  patterns with generative adversarial networks.
\newblock In: IGARSS 2018-2018 IEEE International Geoscience and Remote Sensing
  Symposium, pp. 2095--2098. IEEE (2018)

\bibitem{arcaute2016cities}
Arcaute, E., Molinero, C., Hatna, E., Murcio, R., Vargas-Ruiz, C., Masucci,
  A.P., Batty, M.: Cities and regions in britain through hierarchical
  percolation.
\newblock Royal Society open science \textbf{3}(4), 150691 (2016).
\newblock \doi{https://doi.org/10.1098/rsos.150691}

\bibitem{barthelemy2008modeling}
Barth{\'e}lemy, M., Flammini, A.: Modeling urban street patterns.
\newblock Physical review letters \textbf{100}(13), 138702 (2008)

\bibitem{buhl2006topological}
Buhl, J., Gautrais, J., Reeves, N., Sol{\'e}, R., Valverde, S., Kuntz, P.,
  Theraulaz, G.: Topological patterns in street networks of self-organized
  urban settlements.
\newblock The European Physical Journal B-Condensed Matter and Complex Systems
  \textbf{49}(4), 513--522 (2006)

\bibitem{cardillo2006structural}
Cardillo, A., Scellato, S., Latora, V., Porta, S.: Structural properties of
  planar graphs of urban street patterns.
\newblock Physical Review E \textbf{73}(6), 066107 (2006)

\bibitem{dangeti2017statistics}
Dangeti, P.: Statistics for machine learning.
\newblock Packt Publishing Ltd (2017)

\bibitem{fukushima1980neocognitron}
Fukushima, K.: Neocognitron: A self-organizing neural network model for a
  mechanism of pattern recognition unaffected by shift in position.
\newblock Biological cybernetics \textbf{36}(4), 193--202 (1980)

\bibitem{haklay2008openstreetmap}
Haklay, M., Weber, P.: Openstreetmap: User-generated street maps.
\newblock IEEE Pervasive Computing \textbf{7}(4), 12--18 (2008)

\bibitem{kingma2013auto}
Kingma, D.P., Welling, M.: Auto-encoding variational bayes.
\newblock arXiv preprint arXiv:1312.6114  (2013)

\bibitem{krizhevsky2012imagenet}
Krizhevsky, A., Sutskever, I., Hinton, G.E.: Imagenet classification with deep
  convolutional neural networks.
\newblock In: Advances in neural information processing systems, pp. 1097--1105
  (2012)

\bibitem{lecun1995convolutional}
LeCun, Y., Bengio, Y., et~al.: Convolutional networks for images, speech, and
  time series.
\newblock The handbook of brain theory and neural networks \textbf{3361}(10),
  1995 (1995)

\bibitem{lecun1990handwritten}
LeCun, Y., Boser, B.E., Denker, J.S., Henderson, D., Howard, R.E., Hubbard,
  W.E., Jackel, L.D.: Handwritten digit recognition with a back-propagation
  network.
\newblock In: Advances in neural information processing systems, pp. 396--404
  (1990)

\bibitem{lundberg2017unified}
Lundberg, S.M., Lee, S.I.: A unified approach to interpreting model
  predictions.
\newblock In: Advances in Neural Information Processing Systems, pp. 4765--4774
  (2017)

\bibitem{maaten2008visualizing}
Maaten, L.v.d., Hinton, G.: Visualizing data using t-sne.
\newblock Journal of machine learning research \textbf{9}(Nov), 2579--2605
  (2008)

\bibitem{masucci2009random}
Masucci, A.P., Smith, D., Crooks, A., Batty, M.: Random planar graphs and the
  london street network.
\newblock The European Physical Journal B \textbf{71}(2), 259--271 (2009)

\bibitem{moosavi2017urban}
Moosavi, V.: Urban morphology meets deep learning: Exploring urban forms in one
  million cities, town and villages across the planet.
\newblock arXiv preprint arXiv:1709.02939  (2017)

\bibitem{newman2010networks}
Newman, M.: Networks: an introduction.
\newblock Oxford university press (2010)

\bibitem{murcio2015multifrac}
R, M., P, M.A., E, A., Batty, M.: Multifractal to monofractal evolution of the
  london street network.
\newblock Phys Rev E \textbf{92}(6), 2130 (2015).
\newblock \doi{https://doi.org/10.1103/PhysRevE.92.062130}

\bibitem{lime}
Ribeiro, M.T., Singh, S., Guestrin, C.: "why should {I} trust you?": Explaining
  the predictions of any classifier.
\newblock In: Proceedings of the 22nd {ACM} {SIGKDD} International Conference
  on Knowledge Discovery and Data Mining, San Francisco, CA, USA, August 13-17,
  2016, pp. 1135--1144 (2016)

\bibitem{shrikumar2017learning}
Shrikumar, A., Greenside, P., Kundaje, A.: Learning important features through
  propagating activation differences.
\newblock In: Proceedings of the 34th International Conference on Machine
  Learning-Volume 70, pp. 3145--3153. JMLR. org (2017)

\bibitem{strano2013urban}
Strano, E., Viana, M., da~Fontoura~Costa, L., Cardillo, A., Porta, S., Latora,
  V.: Urban street networks, a comparative analysis of ten european cities.
\newblock Environment and Planning B: Planning and Design \textbf{40}(6),
  1071--1086 (2013)

\bibitem{witten2016data}
Witten, I.H., Frank, E., Hall, M.A., Pal, C.J.: Data Mining: Practical machine
  learning tools and techniques.
\newblock Morgan Kaufmann (2016)

\bibitem{wu2016quantitative}
Wu, Y., Burda, Y., Salakhutdinov, R., Grosse, R.: On the quantitative analysis
  of decoder-based generative models.
\newblock arXiv preprint arXiv:1611.04273  (2016)

\end{thebibliography}

%
%

\end{document}